\documentclass[aps,pra,showpacs,twocolumn]{revtex4}
\usepackage{hyperref}
\usepackage{amsmath}
\usepackage{amssymb}
\usepackage{mathrsfs} 

\usepackage{graphicx}
\usepackage{xcolor}
\newcommand\mb{\boldsymbol}

\begin{document}

\title{Minkowski tensor in electrodynamics of moving media and three rules for construction of the physical tensors in Einstein's special relativity}
\author{Changbiao Wang }
\email{changbiao_wang@yahoo.com}
\affiliation{ShangGang Group, 10 Dover Road, North Haven, CT 06473, USA}

\begin{abstract}
Minkowski applied Einstein's principle of relativity to moving media and developed electrodynamics of moving media.  Like Einstein introduced the EM field-strength tensor  $F^{\mu\nu}$  for electric field $\mathbf{E}$ and magnetic induction $\mathbf{B}$, Minkowski introduced another EM field-strength tensor $G^{\mu\nu}$  for the electric displacement $\mathbf{D}$ and magnetic field $\mathbf{H}$; thus leading to Minkowski tensor.  Recently, Partanen and Tulkki criticize that Minkowski tensor contradicts special relativity, and proposed a mass-polariton stress-energy-momentum (MP SEM) tensor to replace Minkowski tensor.  In this paper, based on a careful analysis of previous literature on this topic, (i) I reasonably argue that Minkowski tensor is a covariant combination of two EM field-strength tensors, and thus all the physical results obtained from Minkowski tensor are already embodied in the two EM field-strength tensors; (ii) I propose three rules for covariantly and self-consistently constructing the physical tensors in Einstein's special relativity, with Minkowski tensor following all the rules while both Abraham tensor and the MP SEM tensor not.  Finally, by enumerating a specific example I show that the Lorentz covariance of a tensor provides no guarantee of the consistency of the tensor with the principle of relativity.
\end{abstract}

\maketitle 

\section{Introduction}
\label{s1}
As is well-known, Minkowski tensor is the stress-energy-momentum tensor in Minkowski's electrodynamics of moving media.  In their work \cite{r1}, Partanen and Tulkki criticize that Minkowski tensor contradicts Einstein's special relativity, and in order to solve this problem, the authors proposed a new tensor, called mass-polariton stress-energy-momentum (MP SEM) tensor. 
 
In this paper, I would like to indicate that Partanen-Tulkki criticism is not physically self-consistent.  I also provide a mathematical proof that the MP SEM tensor proposed by Partanen and Tulkki does not fulfill Lorentz transformation.  During this critical analysis, I reasonably argue that Minkowski tensor is a covariant combination of two EM field-strength tensors, and consequently, all the physical results obtained from Minkowski tensor are already included in the two EM field-strength tensors.  I propose a set of rules for covariantly and self-consistently constructing the physical tensors in special relativity; Minkowski tensor follows all the rules, while both Abraham tensor and the MP SEM tensor do not. Finally, I enumerate a specific example to show that, the Lorentz covariance of a tensor does not guarantee that the tensor will fulfill the principle of relativity.

\section{Non-self-consistent criticism of Minkowski tensor}
\label{s2}
In a recent paper by Partanen and Tulkki \cite{r1}, the authors claim: (i) the stress-energy-momentum (SEM) tensor of the mass-polariton (MP) is Lorentz covariant; (ii) it fulfills the Lorentz transformation of EM field-strength tensors; (iii) Minkowski SEM tensor contradicts the special theory of relativity.

Partanen and Tulkki highlight that they have particularly written their paper for nonexpert readers.  However as a nonexpert reader, I found that their claim (i) and claim (iii) are not logically compatible.  That is to say, Partanen-Tulkki criticism of Minkowski tensor is not physically self-consistent.  My argument is given below.  

In Partanen-Tulkki paper \cite{r1}, the SEM tensor of the MP is defined by Eq.\,(14), where the EM fields $\mathbf{E}$, $\mathbf{B}$, $\mathbf{D}$, and $\mathbf{H}$ are claimed to satisfy the Lorentz transformation given by Eqs.\,(32)-(35). 

It should be emphasized that the Lorentz transformation Eqs.\,(32)-(35) result from the transformation of two EM field-strength tensors (confer: footnote 7 of Ref.\cite{r2}), which offer the assurance of the symmetry in form of Maxwell equations under Lorentz transformations; namely in all inertial frames under Lorentz transformation, Maxwell equations have the same mathematical form and all EM field quantities appearing in the equations have the same physical definitions, usually referred to as Lorentz covariance/invariance of Maxwell equations.  Thus the claimed Lorentz covariance of Partanen-Tulkki MP SEM tensor is based on the physical validity of the two EM field-strength tensors.  In other words, Partanen and Tulkki agree with the physical validity of EM field-strength tensors --- the EM field-strength tensors follow Einstein's special relativity.

In Partanen-Tulkki paper \cite{r1}, the Minkowski SEM tensor is given by Eq.\,(50), which can be written in a \emph{covariant form}, given by 
\begin{equation}
T_{\mathrm{M}}^{\mu\nu}=g^{\mu\sigma}G_{\sigma\lambda}F^{\lambda\nu}+\frac{1}{4}g^{\mu\nu}G_{\sigma\lambda}F^{\sigma\lambda},
\label{eq1}
\end{equation}
where the Minkowski metric is given by $g^{\mu\nu}=g_{\mu\nu}=\mathrm{diag}(-1,-1,-1,+1)$  in terms of the practice used in footnote 7 of Ref.\cite{r2}; $F^{\mu\nu}$ and $G^{\mu\nu}$ are the EM field-strength tensors, given by
\begin{equation*}
F^{\mu\nu}=\left(\begin{array}{cccc} 0 & -B_z & B_y & E_x/c \\ B_z & 0 & -B_x & E_y/c \\ -B_y & B_x & 0 & E_z/c \\ -E_x/c & -E_y/c & -E_z/c & 0 \end{array}\right),~~~~~~~~~~~
\end{equation*}
\begin{equation*}
G_{\mu\nu}=g_{\mu\sigma}G^{\sigma\lambda}g_{\lambda\nu}=\left(\begin{array}{cccc} 0 & -H_z & H_y & -D_xc \\ H_z & 0 & -H_x & -D_yc \\ -H_y & H_x & 0 & -D_zc \\ D_xc & D_yc & D_zc & 0 \end{array}\right),
\end{equation*}
with $c$ the speed of light in free space.

From Eq.\,(\ref{eq1}) it is clearly seen that, (i) all the physical results obtained from Minkowski SEM tensor are already embodied in the two EM field-strength tensors, and (ii) the physical validity of Minkowski SEM tensor is guaranteed by the physical validity of the two EM field-strength tensors  $F^{\mu\nu}$ and $G^{\mu\nu}$.

Partanen and Tulkki argue that ``the Minkowski SEM theory [tensor] is in contradiction with the fundamental principles of the STR [special theory of relativity]''  \cite{r1}.  Thus according to Eq.\,(\ref{eq1}), Partanen-Tulkki argument means that the EM field-strength tensors are ``in contradiction with the fundamental principles of the STR'' --- challenging the physical validity of EM field-strength tensors.

\emph{Conclusion.} On one hand, Partanen and Tulkki \emph{agree with} the physical validity of the EM field-strength tensors when proving the claimed Lorentz covariance of their MP SEM tensor, but on the other hand, they \emph{challenge} the validity by negating Minkowski SEM tensor. From this, I conclude that Partanen-Tulkki criticism of Minkowski tensor is not physically self-consistent. 

\section{Is the Lorentz covariance of the MP SEM tensor Eq.\,(14) proved?}
\label{s3}
The construction of a physical tensor is different from the construction of a mathematical tensor.  The former has more limitations.  Generally speaking, there are three rules to be satisfied for a new physical tensor.  
\begin{description}
\item[Rule (1)] A constructed new physical tensor has a symmetric/covariant/invariant/frame-independent definition,  so that the ``invariance of physical definitions'' is satisfied, required by the principle of relativity \cite{r2}.
\item[Rule (2)] It follows Lorentz transformation, required by the attribute of tensors in mathematics.
\item[Rule (3)] Mathematically and physically, it does not contradict well-established basic physical tensors, such as the phase invariant (zeroth rank tensor), time-space four-vector and wave four-vector (both first rank tensors), EM field-strength tensors (second rank tensors), and so on. This rule is required by the self-consistency of physics theories.
\end{description}   
Unfortunately, some of the above three rules is often neglected in the community so that a constructed new tensor is not a \emph{real} physical tensor; a typical example is the Abraham tensor \cite{r3,r4}, which follows Rule (1), but breaks Rule (2) or Rule (3): if Rule (3) is followed, then Rule (2) is broken; if Rule (2) is followed, then Rule (3) is broken.

In Partanen-Tulkki paper \cite{r1},  the MP SEM tensor Eq.\,(14) is not expressed in a \emph{covariant form}, and it is not easy to judge whether the MP SEM tensor is a real Lorentz covariant physical tensor; for example, does it follow Lorentz transformation?  In other words, does Rule (2) hold for the MP SEM tensor?

To answer the above challenging question effectively and convincingly, I managed to find a mathematical proof that the MP SEM tensor does not follow Lorentz transformation, breaking Rule (2).  The proof is given below.

To understand Partanen-Tulkki MP theory, first we have to find out what basic assumptions are used, and then we can check if the assumptions are justified.

In Partanen-Tulkki theory \cite{r1}, the MP SEM tensor is defined as $\mathbf{T}_{\mathrm{MP}}=\mathbf{T}_{\mathrm{field}}+\mathbf{T}_{\mathrm{MDW}}$, where  $\mathbf{T}_{\mathrm{field}}$ is Abraham EM tensor \cite{r3}, and $\mathbf{T}_{\mathrm{MDW}}$  is the mass density wave (MDW) tensor.  By re-combination of  $\mathbf{T}_{\mathrm{field}}$  and  $\mathbf{T}_{\mathrm{MDW}}$, the MP SEM tensor Eq.\,(14) in \cite{r1} can be written (in a mixed-expression way for clarity) as 
\begin{align}
\mathbf{T}_{\mathrm{MP}}=\mathbf{T}_{\mathrm{M}}&+\left(\frac{\rho_{\mathrm{MDW}}}{\gamma_{\mathrm{a}} \gamma_1}\right)U^{\alpha}_{\mathrm{a}}U^{\beta}_1  \notag  \\
~~
\notag  \\
&+\left[\begin{array}{cc} 
0 & \mathbf{0}~ \vspace{2mm} \\
~(\mathbf{E}\times\mathbf{H}/c+\rho_{\mathrm{MDW}}\mathbf{v}_1c) \\
- (c\mathbf{D}\times\mathbf{B}+\rho_{\mathrm{MDW}}\mathbf{v}_{\mathrm{a}}c) & \mathbf{0}~ \end{array} \right],
\label{eq2}
\end{align} \\
which is identical to the first expression of Eq.\,(51) in \cite{r5}, and where $\mathbf{T}_{\mathrm{M}}$  is the well-known Lorentz covariant Minkowski tensor,  $U^{\alpha}_{\mathrm{a}}=\gamma_{\mathrm{a}}(c,\mathbf{v}_{\mathrm{a}})$  is the atomic four-velocity,  $\mathbf{v}_{\mathrm{a}}$ is the atomic velocity, $\gamma_{\mathrm{a}}=(1-\mb{\beta}_{\mathrm{a}}^2)^{-1/2}$, and $\mb{\beta}_{\mathrm{a}}=\mathbf{v}_{\mathrm{a}}/c$; $U^{\alpha}_{\mathrm{1}}=\gamma_1(c,\mathbf{v}_{\mathrm{1}})$ is the four-velocity of light,    $\mathbf{v}_{\mathrm{1}}$ is the velocity of light, $\gamma_{\mathrm{1}}=(1-\mb{\beta}_{\mathrm{1}}^2)^{-1/2}$, and  $\mb{\beta}_{\mathrm{1}}=\mathbf{v}_{\mathrm{1}}/c$;  $\rho_{\mathrm{MDW}}$ is the excess mass density of atoms in the MDW. Note that $U^{\alpha}_{\mathrm{a}}U^{\beta}_1$  is a four-tensor which follows Lorentz transformation.

To make $\mathbf{T}_{\mathrm{MP}}$  fulfill Lorentz transformation, Partanen and Tulkki implicitly introduced two basic assumptions: \\
\begin{align}
~~~\textbf{Assum-(i)}~~~~~~~~~~\frac{\rho_{\mathrm{MDW}}}{\gamma_{\mathrm{a}} \gamma_1}&=\frac{\rho_{\mathrm{a}}-\rho_{\mathrm{0}}}{\gamma_{\mathrm{a}} \gamma_1}  \notag \\
&=\mathrm{Lorentz~invariant}, 
\label{eq3} 
\\
~~
\notag   
\\ 
~~~\textbf{Assum-(ii)}~~~~~\mathbf{E}\times\mathbf{H}/c^2&+\rho_{\mathrm{MDW}}\mathbf{v}_1 \notag \\
&= \mathbf{D}\times\mathbf{B}+\rho_{\mathrm{MDW}}\mathbf{v}_{\mathrm{a}},
\label{eq4}
\end{align} \\
where $\rho_{\mathrm{a}}$  is the actual mass density of atoms, and $\rho_{\mathrm{0}}$  is the atomic mass density in the absence of the MDW, and which hold in all inertial frames. 

However it should be noted that in their paper \cite{r1}, Partanen and Tulkki did not provide the definition for the atomic mass density $\rho_{\mathrm{a}}$  or  $\rho_{\mathrm{0}}$ in a \emph{general} inertial frame; specifically, the authors did not define the relation between the density in a general frame and its proper density in the atom-rest frame. In a recent complementary paper \cite{r5}, Partanen and Tulkki provided two definitions for the mass density  in a general frame: (i) $\rho_{\mathrm{a}}/\gamma_{\mathrm{a}}$  = proper density (invariant), and (ii) $\rho_{\mathrm{a}}/\gamma^2_{\mathrm{a}}$  = proper density (invariant).  The \emph{one} gamma-factor in definition (i) is required by the definition of ``Lagrangian momentum density four-vector'', given by  $(\rho_{\mathrm{a}}/\gamma_{\mathrm{a}})U^{\alpha}_{\mathrm{a}}=\rho_{\mathrm{a}}(c,\mathbf{v}_{\mathrm{a}})$, while the \emph{two} gamma-factors in definition (ii) are argued to ``originate from the Lorentz contraction and the kinetic energy of'' the proper density.  The two definitions have the same proper density, but they are not compatible, because $\rho_{\mathrm{a}}(c,\mathbf{v}_{\mathrm{a}})$  is a four-vector under definition (i), while it  is not a four-vector under definition (ii).

With Assum-(ii) considered, from Eq.\,(\ref{eq2}) the MP SEM tensor can be rewritten as
\begin{equation}
\mathbf{T}_{\mathrm{MP}}=\mathbf{T}_{\mathrm{M}}+\left(\frac{\rho_{\mathrm{MDW}}}{\gamma_{\mathrm{a}}\gamma_1}\right)U^{\alpha}_{\mathrm{a}}U^{\beta}_1,
\label{eq5}
\end{equation}
which is identical to the second expression of Eq.\,(51) in \cite{r5}.  Now Assum-(i) can be directly recognized from the fact that $g_{\alpha\beta}T_{\mathrm{MP}}^{\alpha\beta}=g_{\alpha\beta}T_{\mathrm{M}}^{\alpha\beta}+[\rho_{\mathrm{MDW}}/(\gamma_{\mathrm{a}}\gamma_1)]g_{\alpha\beta}U^{\alpha}_{\mathrm{a}}U^{\beta}_1$  must be a Lorentz scalar if $T_{\mathrm{MP}}^{\alpha\beta}$  is a real Lorentz tensor, where Minkowski metric is given by $g_{\alpha\beta}=\mathrm{diag}(+1,-1,-1,-1)$, with $\alpha, \beta=0,1,2,3$  in terms of Partanen-Tulkki practice.   Note: $g_{\alpha\beta}U^{\alpha}_{\mathrm{a}}U^{\beta}_1=\gamma_{\mathrm{a}}\gamma_{\mathrm{1}}(c^2-\mathbf{v}_{\mathrm{a}}\cdot\mathbf{v}_{\mathrm{1}})\ne 0$ holds. Since  $g_{\alpha\beta}T_{\mathrm{MP}}^{\alpha\beta}$,  $g_{\alpha\beta}T_{\mathrm{M}}^{\alpha\beta}$, and $g_{\alpha\beta}U^{\alpha}_{\mathrm{a}}U^{\beta}_1$  are all scalars,  $\rho_{\mathrm{MDW}}/(\gamma_{\mathrm{a}}\gamma_1)$ must also be a scalar.

Thus Assum-(i) is a \emph{necessary condition} for both Eq.\,(\ref{eq2}) and Eq.\,(\ref{eq5}) to follow Lorentz transformation, because Eq.\,(\ref{eq2}) and Eq.\,(\ref{eq5}) are equivalent under Assum-(ii).

Unfortunately, Assum-(i) is not justified when a mass density wave (MDW) exists, as proved in  Appendix \ref{appa} for the definition $\rho_{\mathrm{a}}/\gamma_{\mathrm{a}}$ = proper density invariant and Appendix \ref{appb} for the definition $\rho_{\mathrm{a}}/\gamma^2_{\mathrm{a}}$  = proper density invariant, respectively; thus the MP SEM tensor does not follow Lorentz transformation, breaking Rule (2).  

Assums-(i) and (ii) are, respectively, implicitly included in Eqs.\,(36) and (51) of the paper by Partanen and Tulkki \cite{r1}.  It is worthwhile to point out that Eq.\,(36) is actually equivalent to Assum-(i) although they have quite different mathematical forms.  Thus Partanen-Tulkki Eq.\,(36) \cite{r1} is not justified either. A proof of the equivalency between Partanen-Tulkki Eq.\,(36) and Assum-(i) is shown in Appendix \ref{appc}.

\emph{Conclusion.}  I have generally proved in the limit of Partanen-Tulkki theory that the MP SEM tensor cannot fulfill Lorentz transformation, breaking Rule (2).   From this I can conclude that it is not true for Partanen-Tulkki claim that ``we have proved the Lorentz covariance of the MP theory of light'' \cite{r1}.  

\section{Remarks on MP SEM tensor and proposed three rules}
\label{s4}	
Three years after Einstein developed special relativity, Minkowski applied Einstein's principle of relativity to moving media, arguing for invariance of Maxwell equations, and developed electrodynamics of moving media \cite{r6}.  Like Einstein introduced the EM field-strength tensor $F^{\mu\nu}$  to generate the Lorentz transformation of $\mathbf{E}$ and $\mathbf{B}$ \cite{r7}, Minkowski introduced another EM field-strength tensor $G^{\mu\nu}$ to generate the Lorentz transformation of $\mathbf{D}$ and $\mathbf{H}$ \cite[p.\,557]{r8}; thus leading to Minkowski tensor, as shown in Eq.\,(\ref{eq1}). 

Partanen and Tulkki argue that Minkowski tensor is in contradiction with the fundamental principles of Einstein's special relativity \cite{r1}, and proposed MP SEM tensor $\mathbf{T}_{\mathrm{MP}}$ to replace Minkowski tensor, as shown in Eq.\,(\ref{eq5}).  Thus a basic question results: Does  $\mathbf{T}_{\mathrm{MP}}$  fulfill Lorentz transformation?  Actually, this question belongs to a fundamental issue in special relativity: How to covariantly and self-consistently construct a physical tensor.  In this paper, I proposed three rules to solve this issue based on the analysis of Partanen-Tulkki work \cite{r1}.

I would like to emphasize that, whether the MP SEM tensor fulfills Lorentz transformation is the core issue of the whole Partanen-Tulkki theory.  For example, Partanen and Tulkki claim in a recent Comment \cite{r9} that they have proved the \emph{divergence-free} of their MP SEM tensor, namely  $\partial_{\nu}T^{\mu\nu}_{\mathrm{MP}}=0$.  However Partanen-Tulkki proof is not valid, because $T^{\mu\nu}_{\mathrm{MP}}$  does not fulfill Lorentz transformation so that the holding of $\partial_{\nu}T^{\mu\nu}_{\mathrm{MP}}=0$  in the \emph{laboratory frame} (namely the medium-rest frame) is no guarantee of the holding of $\partial_{\nu}T^{\mu\nu}_{\mathrm{MP}}=0$  in a \emph{general frame}.  Unfortunately, Brevik endorsed Partanen-Tulkki proof in his Reply \cite{r10}.

Similarly, if $T^{\mu\nu}_{\mathrm{MP}}$  does not fulfill Lorentz transformation, then the holding of   $T^{\mu\nu}_{\mathrm{MP}}=T^{\nu\mu}_{\mathrm{MP}}$ (symmetry) in the medium-rest frame is no guarantee of the holding of  $T^{\mu\nu}_{\mathrm{MP}}=T^{\nu\mu}_{\mathrm{MP}}$ (symmetry) in a general frame either.

In addition, I also would like to emphasize the significance of the proposed three rules for constructing physical tensors covariantly and self-consistently.  Rule (1) and Rule (2) constitute the definition of covariance for a physical tensor, as shown in footnote 7 of Ref.\,\cite{r2}, while Rule (3) provides the assurance of self-consistency for the tensor. 

As shown in Eq.\,(\ref{eq1}), Minkowski tensor is a covariant combination of EM field-strength tensors, and it satisfy all the three rules.  Abraham tensor cannot follow Lorentz transformation, because EM field-strength tensors \emph{must} follow Lorentz transformation to keep the invariance of Maxwell equations, as shown in \cite{r4,r11}; the MP SEM tensor $T^{\mu\nu}_{\mathrm{MP}}$  cannot either, because  $\rho_{\mathrm{a}}$,  $\gamma_{\mathrm{a}}$,  $\rho_{\mathrm{0}}$, and $\gamma_{\mathrm{1}}$  in $(\rho_{\mathrm{a}}-\rho_{\mathrm{0}})/(\gamma_{\mathrm{a}}\gamma_{\mathrm{1}})$  have to fulfill their own Lorentz transformations, as shown in Appendices \ref{appa} and \ref{appb}.  Thus both Abraham tensor and MP SEM tensor break Rule (2).

One might argue: If a tensor satisfies Rules (1) and (2), then it is Lorentz covariant, while a Lorentz covariant tensor certainly follows the principle of relativity; thus Rule (3) must be redundant.  However this is not true.  To understand this, a specific example is given below.
 
As we know, time dilation, Lorentz contraction, and relativity of simultaneity are the basic results from the time-space four-vector  $\mathrm{d}X^{\mu}=(\mathrm{d}\mathbf{x},\mathrm{d}ct)$ and they constitute entry-level knowledge of special relativity.  Based on the time-space four-vector, the four-momentum for a massive particle is defined by 
\begin{align}
m_0\frac{\mathrm{d}X^{\mu}}{\mathrm{d}\tau}&=m_0\gamma(\mathbf{v},c) \notag \\
&=m\,(\mathbf{v},c)=\left(m\mathbf{v},\frac{mc^2}{c}\right),
\label{eq6}
\end{align} 
where $m_0$  is the particle's rest mass or proper mass invariant, $\tau$  is the proper time, and $\mathrm{d}X^{\mu}/\mathrm{d}\tau=\gamma(\mathbf{v},c)$ is the four-velocity, with $\mathbf{v}=\mathrm{d}\mathbf{x}/\mathrm{d}t$  the particle's velocity and $\gamma=(1-\mathbf{v}^2/c^2)^{-1/2}$.  

In Eq.\,(\ref{eq6}), $m=\gamma m_0$  is defined as the particle's relativistic mass, increasing with the increase of the particle's velocity $\mathbf{v}$.  Thus the Lorentz covariant four-momentum $m_0\gamma(\mathbf{v},c)$  has no contradiction with the time-space four-vector  $\mathrm{d}X^{\mu}=(\mathrm{d}\mathbf{x},\mathrm{d}ct)$, satisfying all the three rules.

Similar to the four-momentum $m_0\gamma(\mathbf{v},c)$, one may construct a differential-element four-vector, given by 
\begin{equation}
\mathrm{d}^3x_0\,\gamma(\mathbf{\mb{\beta}},1)=\mathrm{d}^3x\,(\mathbf{\mb{\beta}},1),
\label{eq7}
\end{equation}
where $\mathrm{d}^3x_0$  is the proper differential-element volume, $\mathbf{\mb{\beta}}=\mathbf{v}/c$ is the normalized velocity of the element, and $\gamma(\mathbf{\mb{\beta}},1)=\gamma(\mathbf{v},c)/c$ is the normalized four-velocity.  Apparently, $\mathrm{d}^3x_0\,\gamma(\mathbf{\mb{\beta}},1)$  is also Lorentz covariant \cite[p.\,757]{r8}, namely satisfying Rules (1) and (2).

From Eq.\,(\ref{eq7}), the relativistic differential-element volume is given by
\begin{equation}
\mathrm{d}^3x=\gamma\,\mathrm{d}^3x_0~~\left[\begin{array}{c}
\textrm{from differential-element} \\
\textrm{four-vector} \end{array}\right],	
\label{eq8}
\end{equation}
which is increasing with the increase of the element's velocity $\mathbf{v}=\mb{\beta}c$, while according to the effect of Lorentz contraction resulting from the time-space four-vector,   $\mathrm{d}^3x$ is supposed to be equal to  $\mathrm{d}^3x_0/\gamma$, namely \cite{r13} 
\begin{equation}
\mathrm{d}^3x=\frac{\mathrm{d}^3x_0}{\gamma}~~\left[\begin{array}{c}
\textrm{from time-space} \\
\textrm{four-vector} \end{array}\right].
\label{eq9}
\end{equation}

We find that Eq.\,(\ref{eq8}) and Eq.\,(\ref{eq9}) are not compatible for $\mathbf{v}=\mb{\beta}c\ne 0$, with the former $\mathrm{d}^3x$ becoming larger while the latter $\mathrm{d}^3x$  becoming smaller.  From this we conclude that, the differential-element four-vector  $\mathrm{d}^3x_0\,\gamma(\mathbf{\mb{\beta}},1)$ physically \emph{contradicts} the effect of Lorentz contraction [resulting from the time-space four-vector  $\mathrm{d}X^{\mu}=(\mathrm{d}\mathbf{x},\mathrm{d}ct)$], breaking Rule (3); and this differential-element four-vector is only a \emph{mathematical} four-vector, instead of a \emph{physical} four-vector.

Obviously, the above example leads us to a conclusion: 
\begin{itemize}
\item The Lorentz covariance of a tensor does not guarantee that the tensor will follow the principle of relativity.
\end{itemize}  
Thus both the covariance and self-consistency are required in constructing physical tensors in Einstein's special relativity. Rules (1) and (2) provide the assurance of covariance, while Rule (3) provides the assurance of self-consistency, as mentioned previously.


\appendix
\section{\label{appa} Contradiction of Assum-(i) against Lorentz transformation for $\rho_{\mathrm{a}}/\gamma_{\mathrm{a}}$ = invariant}
As indicated, in Partane-Tulkki theory there are two definitions for the atomic mass density in a general frame: (i) $\rho_{\mathrm{a}}/\gamma_{\mathrm{a}}$ = Lorentz invariant, and (ii)  $\rho_{\mathrm{a}}/\gamma^2_{\mathrm{a}}$ = Lorentz invariant \cite{r5}. In this appendix, I provide a general mathematical proof for the definition $\rho_{\mathrm{a}}/\gamma_{\mathrm{a}}$ = invariant that  Assum-(i) or Eq.\,(\ref{eq3}) is not true, and the MP SEM tensor in Partanen-Tulkki theory does not follow Lorentz transformation. 

If a tensor  $\mathbf{T}$ satisfies Lorentz transformation at \emph{all} time-space points, then $\mathbf{T}$  is said to be following the Lorentz transformation. If there is \emph{at least one} time-pace point at which $\mathbf{T}$  does not do, then $\mathbf{T}$  is said to be not following the Lorentz transformation. In this paper, whether a tensor follows/satisfies/fulfills Lorentz transformation is in accordance with the above definition.

Because Assum-(i),  $\rho_{\mathrm{MDW}}/(\gamma_{\mathrm{a}}\gamma_1)=(\rho_{\mathrm{a}}-\rho_{0})/(\gamma_{\mathrm{a}}\gamma_1)$ = Lorentz invariant, is not in a \emph{covariant form}, we must check whether it  is really invariant under the Lorentz transformations of $\rho_{\mathrm{a}}$, $\gamma_{\mathrm{a}}$,  $\rho_{\mathrm{0}}$, and  $\gamma_{\mathrm{1}}$. 

Suppose that $XYZ$  and $X'Y'Z'$  are two general inertial frames, with $X'Y'Z'$  moving at $\mb{\beta}c$  with respect to $XYZ$.  According to Partanen-Tulkki MP theory \cite{r1}, we have the following four-vectors. 
\begin{enumerate}
\item $(\rho_{\mathrm{a}}/\gamma_{\mathrm{a}})U^{\alpha}_{\mathrm{a}}=\rho_{\mathrm{a}}(c,\mb{\beta}_{\mathrm{a}}c)$  with $\mb{\beta}_{\mathrm{a}}=\mathbf{v}_{\mathrm{a}}/c$  and $\gamma_{\mathrm{a}}=(1-\mb{\beta}_{\mathrm{a}}^2)^{-1/2}$  is the actual atomic momentum density four-vector \cite{r5}; 

\item $U^{\alpha}_{\mathrm{a}}=\gamma_{\mathrm{a}}(c,\mb{\beta}_{\mathrm{a}}c)$  is the atomic velocity four-vector; 

\item $(\rho_{\mathrm{0}}/\gamma_{\mathrm{0}})U^{\alpha}_{\mathrm{0}}=\rho_{\mathrm{0}}(c,\mb{\beta}_{\mathrm{0}}c)$  with $\mb{\beta}_{\mathrm{0}}=\mathbf{v}_{\mathrm{0}}/c$, $\gamma_{\mathrm{0}}=(1-\mb{\beta}_{\mathrm{0}}^2)^{-1/2}$, and $U^{\alpha}_{\mathrm{0}}=\gamma_{\mathrm{0}}(c,\mb{\beta}_{\mathrm{0}}c)$ is the atomic momentum density four-vector in the absence of the mass density wave (MDW); 

\item  $U^{\alpha}_{\mathrm{1}}=\gamma_{\mathrm{1}}(c,\mb{\beta}_{\mathrm{1}}c)$  with $\mb{\beta}_{\mathrm{1}}=\mathbf{v}_{\mathrm{1}}/c$  and $\gamma_{\mathrm{1}}=(1-\mb{\beta}_{\mathrm{1}}^2)^{-1/2}$ is the velocity four-vector of light.  
\end{enumerate}

According to Partanen-Tulkki Assum-(i), we have
\begin{equation}
\frac{\rho_{\mathrm{a}}-\rho_{0}}{\gamma_{\mathrm{a}}\gamma_1}=\frac{\rho'_{\mathrm{a}}-\rho'_{0}}{\gamma'_{\mathrm{a}}\gamma'_1}=\mathrm{Lorentz~ invariant},
\label{eqa1}
\end{equation} \\
\noindent which is a \emph{necessary condition} for the MP SEM tensor to fulfill Lorentz transformation.  

From $\rho_{\mathrm{a}}(c,\mb{\beta}_{\mathrm{a}}c)$, $\gamma_{\mathrm{a}}(c,\mb{\beta}_{\mathrm{a}}c)$,  $\rho_{\mathrm{0}}(c,\mb{\beta}_{\mathrm{0}}c)$, and  $\gamma_{\mathrm{1}}(c,\mb{\beta}_{\mathrm{1}}c)$, the Lorentz transformations of  $\rho_{\mathrm{a}}'$, $\gamma_{\mathrm{a}}'$,  $\rho_{\mathrm{0}}'$, and  $\gamma_{\mathrm{1}}'$  are, respectively, given by
\begin{align}
\rho_{\mathrm{a}}'&=\gamma\rho_{\mathrm{a}}(1-\mb{\beta}\cdot\mb{\beta}_{\mathrm{a}}),
\label{eqa2}  \\
\gamma_{\mathrm{a}}'&=\gamma\gamma_{\mathrm{a}}(1-\mb{\beta}\cdot\mb{\beta}_{\mathrm{a}}),
\label{eqa3} \\
\rho_{\mathrm{0}}'&=\gamma\rho_{\mathrm{0}}(1-\mb{\beta}\cdot\mb{\beta}_{\mathrm{0}}),
\label{eqa4} \\
\gamma_{\mathrm{1}}'&=\gamma\gamma_{\mathrm{1}}(1-\mb{\beta}\cdot\mb{\beta}_{\mathrm{1}}),
\label{eqa5}
\end{align}
where $\gamma=(1-\mb{\beta}^2)^{-1/2}$.  Inserting Eqs.\,(\ref{eqa2})-(\ref{eqa5}) into Eq.\,(\ref{eqa1}) yields \\
\begin{equation}
\frac{\rho_{\mathrm{a}}-\rho_{0}}{\gamma_{\mathrm{a}}\gamma_1}
=\frac{\gamma\rho_{\mathrm{a}}(1-\mb{\beta}\cdot\mb{\beta}_{\mathrm{a}}) - \gamma\rho_{\mathrm{0}}(1-\mb{\beta}\cdot\mb{\beta}_{\mathrm{0}})}{[\gamma\gamma_{\mathrm{a}}(1-\mb{\beta}\cdot\mb{\beta}_{\mathrm{a}})]\,[\gamma\gamma_{\mathrm{1}}(1-\mb{\beta}\cdot\mb{\beta}_{\mathrm{1}})]},
\label{eqa6}
\end{equation}
or
\begin{equation}
\rho_{\mathrm{a}}-\rho_{0}=\frac{\rho_{\mathrm{a}}(1-\mb{\beta}\cdot\mb{\beta}_{\mathrm{a}})-\rho_{\mathrm{0}}(1-\mb{\beta}\cdot\mb{\beta}_{\mathrm{0}})}{\gamma(1-\mb{\beta}\cdot\mb{\beta}_{\mathrm{1}})(1-\mb{\beta}\cdot\mb{\beta}_{\mathrm{a}})}.
\label{eqa7}
\end{equation} \\

Given any set of ($\rho_{\mathrm{a}}$, $\rho_{\mathrm{0}}$,  $\mb{\beta}_{\mathrm{a}}$,  $\mb{\beta}_{\mathrm{0}}$, $\mb{\beta}_{\mathrm{1}}$), Eq.\,(\ref{eqa7}) is required to be holding for \emph{any}  $|\mb{\beta}|<1$; however, this is impossible.  To understand this, letting $\rho_{\mathrm{0}}=0$  while $\rho_{\mathrm{a}}\ne 0$ (this is allowed mathematically, because they are independent), we have 
\begin{equation}
\gamma(1-\mb{\beta}\cdot\mb{\beta}_{\mathrm{1}})=1,
\label{eqa8}
\end{equation}
which is required to hold for any  $|\mb{\beta}|<1$.  However when $\mb{\beta}=\mb{\beta}_{\mathrm{1}}\ne 0$  is taken, we have $\gamma(1-\mb{\beta}\cdot\mb{\beta}_{\mathrm{1}})=\gamma_1(1-\mb{\beta}_1\cdot\mb{\beta}_{\mathrm{1}})=1/\gamma_1 < 1$ $\Rightarrow$   Eq.\,(\ref{eqa8}) is not true $\Rightarrow$ Eq.\,(\ref{eqa7}) does not hold  $\Rightarrow$  Eq.\,(\ref{eqa1}) is not true.   Thus under Lorentz transformation, $\rho_{\mathrm{MDW}}/(\gamma_{\mathrm{a}}\gamma_1)=(\rho_{\mathrm{a}}-\rho_{0})/(\gamma_{\mathrm{a}}\gamma_1)$  is not Lorentz invariant.  From this we conclude that the MP SEM tensor does not satisfy Lorentz transformation.

In the above proof, a condition of $\rho_{\mathrm{a}}\ne\rho_{\mathrm{0}}$ plus $\rho_{\mathrm{0}}=0$  is used; the proof is much simplified by setting $\rho_{\mathrm{0}}=0$  and easier to understand. However one might argue that in Partanen-Tulkki physical model,  $\rho_{\mathrm{a}}\ne\rho_{\mathrm{0}}$ is valid in general because of the existence of MDW, but   $\rho_0=0$ does not make sense because of the existence of medium. That is true.  To solve this problem, a general proof is provided below, where only $\rho_{\mathrm{a}}\ne\rho_{\mathrm{0}}$  is used, and some ingenious calculations are employed.

Eq.\,(\ref{eqa7}) can be rewritten as 
\begin{equation}
(\rho_{\mathrm{a}}-\rho_{0})F_1(\mb{\beta})=\rho_{\mathrm{a}}F_2(\mb{\beta})-\rho_{\mathrm{0}}F_3(\mb{\beta}),
\label{eqa9}
\end{equation}
where
\begin{align}
F_1(\mb{\beta})&=\gamma(1-\mb{\beta}\cdot\mb{\beta}_{\mathrm{1}})(1-\mb{\beta}\cdot\mb{\beta}_{\mathrm{a}}), 
\label{eqa10} \\
F_2(\mb{\beta})&=(1-\mb{\beta}\cdot\mb{\beta}_{\mathrm{a}}),
\label{eqa11} \\
F_3(\mb{\beta})&=(1-\mb{\beta}\cdot\mb{\beta}_{\mathrm{0}}).
\label{eqa12}
\end{align} \\
\indent Because $\mb{\beta}$  with $|\mb{\beta}|<1$  is arbitrary, we set $\mb{\beta}=(\beta_x, 0, 0)$   with $-1<\beta_x<1$.  Thus we have \\
\begin{align}
F_1(\mb{\beta})&=F_1(\beta_x)=\frac{(1-\beta_x\beta_{1x})(1-\beta_x\beta_{\mathrm{a}x})}{(1-\beta_x^2)^{1/2}}, 
\label{eqa13} \\
F_2(\mb{\beta})&=F_2(\beta_x)=(1-\beta_x\beta_{\mathrm{a}x}),
\label{eqa14} \\
F_3(\mb{\beta})&=F_3(\beta_x)=(1-\beta_x\beta_{\mathrm{0}x}).
\label{eqa15}
\end{align} 

Carrying out integration over both sides of Eq.\,(\ref{eqa9}) with respect to  $\beta_x$, we have				
\begin{align}
(\rho_{\mathrm{a}}-\rho_{0})&\int_{-a}^{a}F_1(\beta_x)\mathrm{d}\beta_x \notag \\
 =\rho_{\mathrm{a}}&\int_{-a}^{a}F_2(\beta_x)\mathrm{d}\beta_x 
-\rho_{\mathrm{0}}\int_{-a}^{a}F_3(\beta_x)\mathrm{d}\beta_x,
\label{eqa16}
\end{align}
where $a$  with $0\le a<1$  is arbitrary, and
\begin{align}
\int_{-a}^{a}F_1(\beta_x)\mathrm{d}\beta_x&=\beta_{\mathrm{a}x}\beta_{\mathrm{1}x}(-a\sqrt{1-a^2}+\arcsin{a}) \nonumber \\
&~~~+2\arcsin{a},
\label{eqa17} \\
~~
\notag \\
\int_{-a}^{a}F_2(\beta_x)\mathrm{d}\beta_x&=\int_{-a}^{a}F_3(\beta_x)\mathrm{d}\beta_x=2a.
\label{eqa18}
\end{align}
Inserting  Eq.\,(\ref{eqa17}) and Eq.\,(\ref{eqa18}) into Eq.\,(\ref{eqa16}), then with $(\rho_{\mathrm{a}}-\rho_{\mathrm{0}})\ne 0$  taken into account and both sides divided by $(\rho_{\mathrm{a}}-\rho_{\mathrm{0}})$, we have
\begin{align}
\beta_{\mathrm{a}x}\beta_{\mathrm{1}x}(-a\sqrt{1-a^2}&+\arcsin{a})  \nonumber \\
&+2\arcsin{a}=2a,
\label{eqa19}
\end{align}
holding for any $0\le a<1$ , which is a \emph{necessary condition} for the MP SEM tensor to fulfill Lorentz transformation.

For  $a=1/\sqrt{2}$, leading to $\arcsin{(1/\sqrt{2})}=\pi/4$, from Eq.\,(\ref{eqa19}) we have
\begin{equation}
\beta_{\mathrm{a}x}\beta_{\mathrm{1}x}=-\frac{2\pi-4\sqrt{2}}{\pi-2}\approx -0.55.
\label{eqa20}
\end{equation}
For  $a=1/2$, leading to $\arcsin{(1/2)}=\pi/6$, from Eq.\,(\ref{eqa19}) we have
\begin{equation}
\beta_{\mathrm{a}x}\beta_{\mathrm{1}x}=-\frac{4\pi-12}{2\pi-3\sqrt{3}}\approx -0.52.
\label{eqa21}
\end{equation} \\
\indent Apparently, there are no $\beta_{\mathrm{a}x}$ and $\beta_{\mathrm{1}x}$  to satisfy Eq.\,(\ref{eqa20}) and Eq.\,(\ref{eqa21}) at the same time $\Rightarrow$  there are no $\beta_{\mathrm{a}x}$ and $\beta_{\mathrm{1}x}$ to satisfy Eq.\,(\ref{eqa19}) for any  $a$  with $0\le a<1$    $\Rightarrow$   there are no $\mb{\beta}_{\mathrm{a}}$ and $\mb{\beta}_{\mathrm{1}}$  to satisfy Eq.\,(\ref{eqa9}) for any   $\mb{\beta}$ with  $|\mb{\beta}|<1$  $\Rightarrow$   there are no $\mb{\beta}_{\mathrm{a}}$ and $\mb{\beta}_{\mathrm{1}}$  to satisfy Eq.\,(\ref{eqa6}) for any   $\mb{\beta}$ with  $|\mb{\beta}|<1$   $\Rightarrow$  the \emph{necessary condition} Eq.\,(\ref{eqa1}) does not hold for $(\rho_{\mathrm{a}}-\rho_{\mathrm{0}})\ne 0$.  In Partanen-Tulkki theory, at least there is one time-space point where $\rho_{\mathrm{MDW}}=(\rho_{\mathrm{a}}-\rho_{\mathrm{0}})\ne 0$  holds, otherwise no mass density wave exists.  Thus I finish the proof that the MP SEM tensor does not follow Lorentz transformation.

The contradiction of Assum-(i) against Lorentz transformation originates from the fact that $\rho_{\mathrm{a}}$,  $\gamma_{\mathrm{a}}$,  $\rho_{\mathrm{0}}$, and $\gamma_{\mathrm{1}}$  all follow their own Lorentz transformations; when they are inappropriately combined together and forced to satisfy an additional constraint Assum-(i), the contradiction takes place. 

\section{\label{appb} Contradiction of Assum-(i) against Lorentz transformation for $\rho_{\mathrm{a}}/\gamma^2_{\mathrm{a}}$ = invariant}

In Appendix \ref{appa}, I provided a general mathematical proof that Assum-(i) is not true.  In that proof, $\rho_{\mathrm{a}}/\gamma_{\mathrm{a}}$  is taken as a Lorentz invariant because the atomic mass density $\rho_{\mathrm{a}}$  in a general inertial frame is defined through the ``Lagrangian momentum density four-vector''   $(\rho_{\mathrm{a}}/\gamma_{\mathrm{a}})U^{\alpha}_{\mathrm{a}}=\rho_{\mathrm{a}}(c,\mb{\beta}_{\mathrm{a}}c)$ \cite{r5}.  However, Partanen and Tulkki also provided another \emph{non-self-consistent} definition for the atomic mass density $\rho_{\mathrm{a}}$  in a general frame, given by $\rho_{\mathrm{a}}/\gamma^2_{\mathrm{a}}$ = Lorentz invariant \cite{r5}.  In this definition, $\rho_{\mathrm{a}}(c,\mb{\beta}_{\mathrm{a}}c)$  is not a four-vector any more.  (That is why I call it being ``non-self-consistent''.)  In this appendix, I will show that under the definition of $\rho_{\mathrm{a}}/\gamma^2_{\mathrm{a}}$ = invariant, the conclusion obtained in Appendix \ref{appa} is still valid, namely $\rho_{\mathrm{MDW}}/(\gamma_{\mathrm{a}}\gamma_1)=(\rho_{\mathrm{a}}-\rho_{0})/(\gamma_{\mathrm{a}}\gamma_1)$  is not a Lorentz invariant when a mass density wave (MDW) exists. 

Suppose that $XYZ$  and $X'Y'Z'$  are two general inertial frames, with  $X'Y'Z'$  moving at $\mb{\beta}c$  with respect to   $XYZ$.  According to Partanen-Tulkki MP theory \cite{r1,r5}, we have the following Lorentz four-vectors and invariants. 

\begin{enumerate}

\item $U^{\alpha}_{\mathrm{a}}=\gamma_{\mathrm{a}}(c,\mb{\beta}_{\mathrm{a}}c)$  is the actual atomic velocity four-vector,  with $\mb{\beta}_{\mathrm{a}}=\mathbf{v}_{\mathrm{a}}/c$  and $\gamma_{\mathrm{a}}=(1-\mb{\beta}_{\mathrm{a}}^2)^{-1/2}$;

\item  $\rho_{\mathrm{a}}/\gamma^2_{\mathrm{a}}$ = Lorentz invariant, with the actual atomic mass density $\rho_{\mathrm{a}}$ physically equal to the atomic \emph{relativistic} number density in volume multiplied by atomic \emph{relativistic mass};

\item $U^{\alpha}_{\mathrm{0}}=\gamma_{\mathrm{0}}(c,\mb{\beta}_{\mathrm{0}}c)$  is the atomic velocity four-vector in the absence of the MDW, with $\mb{\beta}_{\mathrm{0}}=\mathbf{v}_{\mathrm{0}}/c$ and $\gamma_{\mathrm{0}}=(1-\mb{\beta}_{\mathrm{0}}^2)^{-1/2}$; 

\item  $\rho_{\mathrm{0}}/\gamma^2_{\mathrm{0}}$ = Lorentz invariant, with  $\rho_{\mathrm{0}}$ the atomic mass density in the absence of the MDW;

\item $U^{\alpha}_{\mathrm{1}}=\gamma_{\mathrm{1}}(c,\mb{\beta}_{\mathrm{1}}c)$ is the velocity four-vector of light,  with $\mb{\beta}_{\mathrm{1}}=\mathbf{v}_{\mathrm{1}}/c$  and $\gamma_{\mathrm{1}}=(1-\mb{\beta}_{\mathrm{1}}^2)^{-1/2}$.

\end{enumerate}

According to Partanen-Tulkki Assum-(i), we have
\begin{equation}
\frac{\rho_{\mathrm{a}}-\rho_{0}}{\gamma_{\mathrm{a}}\gamma_1}=\frac{\rho'_{\mathrm{a}}-\rho'_{0}}{\gamma'_{\mathrm{a}}\gamma'_1}=\mathrm{Lorentz~ invariant},
\label{eqb1}
\end{equation} \\
\noindent which is a \emph{necessary condition} for the MP SEM tensor to fulfill Lorentz transformation.  

From $\rho_{\mathrm{a}}/\gamma_{\mathrm{a}}^2$, $\gamma_{\mathrm{a}}(c,\mb{\beta}_{\mathrm{a}}c)$,  $\rho_{\mathrm{0}}/\gamma_{\mathrm{0}}^2$, $\gamma_{\mathrm{0}}(c,\mb{\beta}_{\mathrm{0}}c)$, and  $\gamma_{\mathrm{1}}(c,\mb{\beta}_{\mathrm{1}}c)$, the Lorentz transformations of  $\rho_{\mathrm{a}}'$, $\gamma_{\mathrm{a}}'$,  $\rho_{\mathrm{0}}'$, and  $\gamma_{\mathrm{1}}'$  are, respectively, given by
\begin{align}
\rho_{\mathrm{a}}'&=\rho_{\mathrm{a}}[\gamma(1-\mb{\beta}\cdot\mb{\beta}_{\mathrm{a}})]^2,
\label{eqb2}  \\
\gamma_{\mathrm{a}}'&=\gamma_{\mathrm{a}}\gamma(1-\mb{\beta}\cdot\mb{\beta}_{\mathrm{a}}),
\label{eqb3} \\
\rho_{\mathrm{0}}'&=\rho_{\mathrm{0}}[\gamma(1-\mb{\beta}\cdot\mb{\beta}_{\mathrm{0}})]^2,
\label{eqb4} \\
\gamma_{\mathrm{1}}'&=\gamma_{\mathrm{1}}\gamma(1-\mb{\beta}\cdot\mb{\beta}_{\mathrm{1}}),
\label{eqb5}
\end{align}
where $\gamma=(1-\mb{\beta}^2)^{-1/2}$.  Inserting Eqs.\,(\ref{eqb2})-(\ref{eqb5}) into Eq.\,(\ref{eqb1}) yields
\begin{equation}
\frac{\rho_{\mathrm{a}}-\rho_{0}}{\gamma_{\mathrm{a}}\gamma_1}
=\frac{\rho_{\mathrm{a}}[\gamma(1-\mb{\beta}\cdot\mb{\beta}_{\mathrm{a}})]^2 - \rho_{\mathrm{0}}[\gamma(1-\mb{\beta}\cdot\mb{\beta}_{\mathrm{0}})]^2}{[\gamma_{\mathrm{a}}\gamma(1-\mb{\beta}\cdot\mb{\beta}_{\mathrm{a}})]\,[\gamma_{\mathrm{1}}\gamma(1-\mb{\beta}\cdot\mb{\beta}_{\mathrm{1}})]},~~~~~
\label{eqb6}
\end{equation}
or
\begin{equation}
~~\rho_{\mathrm{a}}-\rho_{0}=
\frac{\rho_{\mathrm{a}}(1-\mb{\beta}\cdot\mb{\beta}_{\mathrm{a}})^2 - \rho_{\mathrm{0}}(1-\mb{\beta}\cdot\mb{\beta}_{\mathrm{0}})^2}{(1-\mb{\beta}\cdot\mb{\beta}_{\mathrm{a}})\,(1-\mb{\beta}\cdot\mb{\beta}_{\mathrm{1}})}.~~~~~~~~
\label{eqb7}
\end{equation}

Given any set of ($\rho_{\mathrm{a}}$, $\rho_{\mathrm{0}}$,  $\mb{\beta}_{\mathrm{a}}$,  $\mb{\beta}_{\mathrm{0}}$, $\mb{\beta}_{\mathrm{1}}$), Eq.\,(\ref{eqb7}) is required to be holding for \emph{any}  $|\mb{\beta}|<1$. Below I will show that a necessary condition for the validity of Eq.\,(\ref{eqb7}) for  $\rho_{\mathrm{0}}\ne 0$  is that  $\mb{\beta}_{\mathrm{a}}=\mb{\beta}_{\mathrm{0}}$  or  $\mathbf{v}_{\mathrm{a}}=\mathbf{v}_{\mathrm{0}}$ holds.

Eq.\,(\ref{eqb7}) can be written as 
\begin{align}
F(\mb{\beta})&:=(\rho_{\mathrm{a}}-\rho_{0})(1-\mb{\beta}\cdot\mb{\beta}_{\mathrm{a}})(1-\mb{\beta}\cdot\mb{\beta}_{\mathrm{1}})
\notag \\
&-\rho_{\mathrm{a}}(1-\mb{\beta}\cdot\mb{\beta}_{\mathrm{a}})^2
+\rho_{\mathrm{0}}(1-\mb{\beta}\cdot\mb{\beta}_{\mathrm{0}})^2=0.
\label{eqb8}
\end{align} 
Eq.\,(\ref{eqb8}) only has the first and second order terms of $\mb{\beta}$.  Thus by recombination, Eq.\,(\ref{eqb8}) can be written as a sum of the first and second order terms, given by
\begin{equation}
F(\mb{\beta})=F_1(\mb{\beta})+F_2(\mb{\beta})=0, 
\label{eqb9}
\end{equation}
where
\begin{align}
F_1(\mb{\beta})=&-(\rho_{\mathrm{a}}-\rho_{0})(\mb{\beta}_{\mathrm{a}}
+\mb{\beta}_{\mathrm{1}})\cdot\mb{\beta}
\notag \\
&+2\rho_{\mathrm{a}}(\mb{\beta}_{\mathrm{a}}\cdot\mb{\beta})-2\rho_{\mathrm{0}}(\mb{\beta}_{\mathrm{0}}\cdot\mb{\beta}),
\label{eqb10}
\end{align}
\begin{align}
F_2(\mb{\beta})=&+(\rho_{\mathrm{a}}-\rho_{0})(\mb{\beta}_{\mathrm{a}}\cdot\mb{\beta})(\mb{\beta}_{\mathrm{1}}\cdot\mb{\beta})
\notag \\
&-\rho_{\mathrm{a}}(\mb{\beta}_{\mathrm{a}}\cdot\mb{\beta})^2+\rho_{\mathrm{0}}(\mb{\beta}_{\mathrm{0}}\cdot\mb{\beta})^2.
\label{eqb11}
\end{align}
Note that Eq.\,(\ref{eqb9}) is an \emph{identity} of $\mb{\beta}$.

Making operation $\nabla_{\mb{\beta}}:=\partial/\partial\mb{\beta}$  on the both sides of Eq.\,(\ref{eqb9}) and then setting  $\mb{\beta}=0$, with $\nabla_{\mb{\beta}}F_2(\mb{\beta})|_{\mb{\beta}=0}=0$  taken into account, we obtain a vector equation 
\begin{equation}
-(\rho_{\mathrm{a}}-\rho_{0})(\mb{\beta}_{\mathrm{a}}
+\mb{\beta}_{\mathrm{1}})+2\rho_{\mathrm{a}}\mb{\beta}_{\mathrm{a}}-2\rho_{\mathrm{0}}\mb{\beta}_{\mathrm{0}}=0.
\label{eqb12}
\end{equation} 

Making operation  $\nabla_{\mb{\beta}}\nabla_{\mb{\beta}}:=(\partial/\partial\mb{\beta})(\partial/\partial\mb{\beta})$  on the both sides of Eq.\,(\ref{eqb9}), with 
$\nabla_{\mb{\beta}}\nabla_{\mb{\beta}}F_1(\mb{\beta})=0$, 
 ~$\nabla_{\mb{\beta}}\nabla_{\mb{\beta}}[(\mb{\beta}_{\mathrm{a}}\cdot\mb{\beta})(\mb{\beta}_{\mathrm{1}}\cdot\mb{\beta})]=\mb{\beta}_{\mathrm{1}}\mb{\beta}_{\mathrm{a}}+\mb{\beta}_{\mathrm{a}}\mb{\beta}_{\mathrm{1}}$, 
 $\nabla_{\mb{\beta}}\nabla_{\mb{\beta}}(\mb{\beta}_{\mathrm{a}}\cdot\mb{\beta})^2
=2\mb{\beta}_{\mathrm{a}}\mb{\beta}_{\mathrm{a}}$,
and $\nabla_{\mb{\beta}}\nabla_{\mb{\beta}}(\mb{\beta}_{\mathrm{0}}\cdot\mb{\beta})^2
=2\mb{\beta}_{\mathrm{0}}\mb{\beta}_{\mathrm{0}}$ taken into account, we obtain a dyadic tensor equation, given by
\begin{align}
(\rho_{\mathrm{a}}-\rho_{0})(\mb{\beta}_{\mathrm{a}}\mb{\beta}_{\mathrm{1}}+\mb{\beta}_{\mathrm{1}}\mb{\beta}_{\mathrm{a}})
&-2\rho_{\mathrm{a}}\mb{\beta}_{\mathrm{a}}\mb{\beta}_{\mathrm{a}}
\notag \\
&+2\rho_{\mathrm{0}}\mb{\beta}_{\mathrm{0}}\mb{\beta}_{\mathrm{0}}=0.
\label{eqb13}
\end{align}

Making a dyadic tensor equation  $\mb{\beta}_{\mathrm{a}}$Eq.\,(\ref{eqb12}) + Eq.\,(\ref{eqb12})$\mb{\beta}_{\mathrm{a}}$, and then making the sum of  $\mb{\beta}_{\mathrm{a}}$Eq.\,(\ref{eqb12}) + Eq.\,(\ref{eqb12})$\mb{\beta}_{\mathrm{a}}$ + Eq.\,(\ref{eqb13}), we obtain
\begin{equation}
\rho_{\mathrm{0}}(\mb{\beta}_{\mathrm{a}}\mb{\beta}_{\mathrm{a}}-\mb{\beta}_{\mathrm{a}}\mb{\beta}_{\mathrm{0}}
-\mb{\beta}_{\mathrm{0}}\mb{\beta}_{\mathrm{a}}+\mb{\beta}_{\mathrm{0}}\mb{\beta}_{\mathrm{0}})=0.
\label{eqb14}
\end{equation}
For  $\rho_{\mathrm{0}}\ne 0$, Eq.\,(\ref{eqb14}) can be written as
\begin{equation}
(\mb{\beta}_{\mathrm{a}}-\mb{\beta}_{\mathrm{0}})(\mb{\beta}_{\mathrm{a}}-\mb{\beta}_{\mathrm{0}})=0.
\label{eqb15}
\end{equation} 

Thus from above  Eq.\,(\ref{eqb15}) we have  $(\mb{\beta}_{\mathrm{a}}-\mb{\beta}_{\mathrm{0}})=0$ holding, namely 
\begin{equation}
\mb{\beta}_{\mathrm{a}}=\mb{\beta}_{\mathrm{0}},
\label{eqb16}
\end{equation} 
which is a \emph{necessary} condition for the holding of Assum-(i) for  $\rho_{\mathrm{0}}\ne 0$.  In Partanen-Tulkki theory,  $\rho_{\mathrm{0}}\ne 0$  holds because of the existence of dielectric material; however, at least there is one time-space point where   $\mb{\beta}_{\mathrm{a}}=\mb{\beta}_{\mathrm{0}}$ or $\mathbf{v}_{\mathrm{a}}=\mathbf{v}_{\mathrm{0}}$ does not hold, otherwise there is no mass density wave existing. (Note that  $\mathbf{v}_{\mathrm{a}}$ is the actual atomic velocity, and $\mathbf{v}_{\mathrm{0}}$  is the atomic velocity in the absence of the mass density wave.)  Thus again, I finish the proof in the limit of Partanen-Tulkki theory that for  $\rho_{\mathrm{a}}/\gamma_{\mathrm{a}}^2$ = invariant, Assum-(i) does not hold either. 

\section{\label{appc} Proof of the equivalency between Partanen-Tulkki Eq.\,(36) and Assum-(i)}
In this appendix, I provide a proof that Eq.\,(36) in Partanen-Tulkki paper \cite{r1} and Assum-(i) are equivalent.

Eq.\,(36) and Eq.\,(38) in Partanen-Tulkki paper \cite{r1} read:
\begin{align}
\rho'_{\mathrm{MDW}}&=\frac{c^2-\mathbf{v}_{\mathrm{1}}\cdot\mathbf{v}}{c^2-(\mathbf{v}\ominus\mathbf{v}_{\mathrm{a}})\cdot\mathbf{v}}\,\rho_{\mathrm{MDW}},
\label{eqc1}
\\
&\textrm{\big(Partanen-Tulkki Eq.\,(36)\big)}
\notag \\
~~~
\notag \\
&~~~~~~~\mathbf{v}'_{\mathrm{a}}=-(\mathbf{v}\ominus\mathbf{v}_{\mathrm{a}}).
\label{eqc2}
\\
&\textrm{\big(Partanen-Tulkki Eq.\,(38)\big)}
\notag
\end{align} 

Inserting Eq.\,(\ref{eqc2}) into Eq.\,(\ref{eqc1}) yields  \vspace{3mm}
\begin{equation}
\rho'_{\mathrm{MDW}}(c^2+\mathbf{v}'_{\mathrm{a}}\cdot\mathbf{v})=\rho_{\mathrm{MDW}}(c^2-\mathbf{v}_{\mathrm{1}}\cdot\mathbf{v}).
\label{eqc3}
\end{equation} \vspace{0.0mm}

$\gamma'_1(c,\mathbf{v}'_1)$  is the four-velocity of light, expressed in the $\mathrm{G}'$  frame, which moves at $\mathbf{v}$  with respect to the $\mathrm{G}$  frame.  According to the Lorentz transformation of  $\gamma'_1(c,\mathbf{v}'_1)$, we have
\begin{equation}
\gamma'_1c=\gamma\left(\gamma_1c -\frac{\mathbf{v}}{c}\cdot\gamma_1\mathbf{v}_1\right)=\frac{\gamma\gamma_1}{c}(c^2-\mathbf{v}\cdot\mathbf{v}_1),
\label{eqc4}
\end{equation}
or
\begin{equation}
(c^2-\mathbf{v}\cdot\mathbf{v}_1)=\frac{\gamma'_1c^2}{\gamma\gamma_1},
\label{eqc5}
\end{equation} \\
where $\gamma=(1-\mathbf{v}^2/c^2)^{-1/2}$.

$\gamma_{\mathrm{a}}(c,\mathbf{v}_{\mathrm{a}})$  is the atomic four-velocity, expressed in the $\mathrm{G}$  frame.  According to the Lorentz transformation of  $\gamma_{\mathrm{a}}(c,\mathbf{v}_{\mathrm{a}})$, we have
\begin{equation}
\gamma_{\mathrm{a}}c=\gamma\left(\gamma'_{\mathrm{a}}c +\frac{\mathbf{v}}{c}\cdot\gamma'_{\mathrm{a}}\mathbf{v}'_{\mathrm{a}}\right)=\frac{\gamma\gamma'_{\mathrm{a}}}{c}(c^2+\mathbf{v}\cdot\mathbf{v}'_{\mathrm{a}}),
\label{eqc6}
\end{equation}
or
\begin{equation}
(c^2+\mathbf{v}\cdot\mathbf{v}'_{\mathrm{a}})=\frac{\gamma_{\mathrm{a}}c^2}{\gamma\gamma'_{\mathrm{a}}}.
\label{eqc7}
\end{equation}

Inserting Eq.\,(\ref{eqc5}) and Eq.\,(\ref{eqc7}) into Eq.\,(\ref{eqc3}) yields
\begin{equation}
\rho'_{\mathrm{MDW}}\frac{\gamma_{\mathrm{a}}c^2}{\gamma\gamma'_{\mathrm{a}}}=\rho_{\mathrm{MDW}}\frac{\gamma'_1c^2}{\gamma\gamma_1},
\label{eqc8}
\end{equation}
or
\begin{equation}
\frac{\rho'_{\mathrm{MDW}}}{\gamma'_{\mathrm{a}}\gamma'_{\mathrm{1}}}=\frac{\rho_{\mathrm{MDW}}}{\gamma_{\mathrm{a}}\gamma_{\mathrm{1}}},
\label{eqc9}
\end{equation}
namely Assum-(i).

From above it is seen that, Partanen-Tulkki Eq.\,(36) = Eq.\,(\ref{eqc1}) and Assum-(i) = Eq.\,(\ref{eqc9})  are indeed equivalent, because of Eq.\,(\ref{eqc1}) $\Longleftrightarrow$  Eq.\,(\ref{eqc3}) $\Longleftrightarrow$ Eq.\,(\ref{eqc8}) $\Longleftrightarrow$  Eq.\,(\ref{eqc9}), with Eq.\,(\ref{eqc5}) and Eq.\,(\ref{eqc7}) taken into account.  In other words, the holding of Partanen-Tulkki Eq.\,(36) is a sufficient and necessary condition for the holding of Assum-(i), and vice versa. 

Thus I finish the proof of the equivalency between Eq.\,(36) of Partanen-Tulkki papper \cite{r1} and Assum-(i). \\ \\ \\ \\


\end{document}